\documentclass[%
,longbibliography%
,reprint%
]{revtex4-1}
\usepackage{graphicx}
\usepackage{amsmath}
\usepackage{fancyhdr}
\newlength{\figsize}
\begin{document}
% \markboth{Andrey V. Panov}{Optical Kerr nonlinearity of disordered all-dielectric resonant high index metasurfaces with negative refraction}
% \catchline{}{}{}{}{}
\title{Optical Kerr nonlinearity of disordered all-dielectric resonant high index metasurfaces with negative refraction}
% \author[1,*]{Andrey V. Panov}
\author{Andrey V. Panov}
%\author{Andrey V. Panov\thanks{e-mail: andrej.panov@gmail.com}}
%\rauthor{A. V. Panov}
%\sodauthor{A. V. Panov}
%\rtitle{Optical Kerr nonlinearity of disordered all-dielectric resonant high index metasurfaces with negative refraction}
%\sodtitle{Optical Kerr nonlinearity of disordered all-dielectric resonant high index metasurfaces with negative refraction}

\affiliation{
% \institute{
% \affil[1]{
%\address{
Institute of Automation and Control Processes,
Far East Branch of Russian Academy of Sciences,
5, Radio st., Vladivostok, 690041, Russian Federation}
% \\ andrej.panov@gmail.com}
%\ead{andrej.panov@gmail.com}
% \mail{e-mail \textsf{andrej.panov@gmail.com}}
 \email{Electronic mail: andrej.panov@gmail.com}
% \pacs{78.67.Pt}{Multilayers; superlattices; photonic structures; metamaterials}
% \pacs{42.65.Hw}{Phase conjugation; photorefractive and Kerr effects}
% \pacs{42.65.An}{Optical susceptibility, hyperpolarizability}
% \PACS{
% {78.67.Pt}{Multilayers; superlattices; photonic structures; metamaterials} \and
% {42.65.Hw}{Phase conjugation; photorefractive and Kerr effects} \and
% {42.65.An}{Optical susceptibility, hyperpolarizability}
% }
% \maketitle

 \begin{abstract}
%\abstract{
The optical Kerr effect of material with negative refraction is
estimated for the first time.
This is done via three-dimensional  finite-difference time-domain (FDTD) simulations of disordered bidisperse metasurfaces consisting of high index (GaP) spheres at the wavelength of 532~nm.
The metasurfaces comprise spherical particles randomly arranged on plane having two sizes close to the magnetic and electric dipole Mie resonances.
The real part of the effective nonlinear refractive index of the metasurfaces is computed in the vicinity of the Mie resonances where the metasurface possesses the negative index of refraction.
The optical Kerr nonlinearity has a peak under the condition for the negative refraction.
Intensity-dependent refractive index of the bidisperse metasurfaces is studied
through concentration transition to the negative refraction state. 
It is shown that the nonlinear Kerr coefficient of the monolayer metasurface has maximum when the effective linear
refractive index is close to zero.
%}
 \end{abstract}

 \doi{10.1134/S0021364020010038}

% \keywords{
% optical Kerr nonlinearity; disordered metasurface; dielectric nanoparticles; negative refraction
% }
%\begin{document}
\maketitle

%\section{Introduction}

In recent years, nonlinear optical properties of all-di\-electric high index metasurfaces have attracted a significant interest of researchers \cite{Smirnova16}.
For example, silicon metasurfaces exhibited enhancement of third harmonic generation by several orders of magnitude compared to the massive material \cite{Shcherbakov14,Yang15}.
The intensity-dependent refractive index  is regularly used in designing all-optical compact swi\-tches.
As demonstrated in Ref.~\cite{Shcherbakov15}, all-optical switching of femtosecond laser pulses passing through flat nanostructure of subwavelength  silicon nanodisks at their magnetic dipolar resonance occurs owing to two-photon absorption being enhanced by a factor of 80 with respect to the unpatterned film.
In Ref%s
.~\cite{%Panov18a,
Panov19}, random monodisperse metasurfaces of gallium phosphide (GaP) spheres near Mie resonances were shown by three-dimensional FDTD modeling to have an optical Kerr effect exceeding by two orders of intensity that of the bulk gallium phosphide. 
Moreover, being single negative metamaterials, the monodisperse metasurfaces with sphere sizes in the vicinity of the Mie resonances reveal the inversion of the sign of the second-order nonlinear refractive index~\cite{Panov19}. 
The possibility of negative refraction by the disordered metasurface consisting of GaP spheres with two radii close to the first magnetic and electric Mie resonances was demonstrated in Ref.~\cite{Panov19a}.
However, until now, the effective Kerr nonlinearity of the negative index metamaterials has not been evaluated. 

Transparent conductive oxide thin films display enhanced nonlinear optical properties at epsilon-near-zero region \cite{Alam16,Caspani16,Carnemolla18}.
It is of importance to investigate the optical nonlinearity of the Mie resonant metasurface at the transition of effective refractive index through zero value.

In this work, the optical Kerr nonlinearity of random metasurfaces having the negative effective refractive index is investigated using three-dimensional FDTD simulations. 
The second-order nonlinear refractive index is also calculated for bidisperse mixtures with various sizes or concentrations of GaP spheres in proximity to the negative refraction regime.

% \section{FDTD simulation details}

The procedure of computation of the real part of the effective nonlinear refractive index of the monolayer nanocomposites is described in depth in Ref.~\cite{Panov18}.
In this procedure, the simulated Gaussian beam falls at normal
incidence on the slab specimen with intensity-dependent index of refraction
$$
 n=n_0+n_2 I,
$$ 
where $n_0$ is the linear refractive index, $n_2$ is the second-order nonlinear refractive index, and $I$ is the intensity of the wave. 
This specimen may be inhomogeneous, for example, consisting of nanoparticles.
Further, the phase change on the axis of the transmitted beam is computed in several points at far distance from the specimen. 
The linear fit of the phase change versus the intensity of the Gaussian beam gives the real part of the effective second-order index of refraction $n_2$. 
The computation of the nonlinear refractive index in several points allows one to estimate the average and standard deviation of $n_2$.
Here, the effective nonlinear refractive index represents the ensemble-averaged optical properties of the specimen. 
Sometimes, it is designated by the term ``equivalent refractive index'' which is used for composite materials with inclusion sizes  being a few tenths of a wavelength.

The simulation parameters in the present study were same as given in Ref.~\cite{Panov19}:
the size of the FDTD computational domain was $4\times 4\times 30$~$\mu$m, the space resolution of the simulations was 5~nm. 
The examined sample was a disordered monolayer comprising the equal numbers of GaP spheres of two radii surrounded by vacuum.
This simplified scheme permits one to exclude the impact of substrate on the metasurface optical properties.
The thickness of the monolayer slab was equal to the largest diameter of the spheres in this monolayer.
Unless otherwise specified, the net number of the particles was 264 on $4\times 4$~$\mu$m area.
At wavelength $\lambda=532$~nm used in these study, the linear refractive index of GaP inclusions $n_{0\,\mathrm{in}}=3.49$ \cite{Aspnes83},
the extinction coefficient $\kappa$ was neglected as it has low value (0.0026) for GaP at $\lambda=532$~nm.
Also, the MEEP (Massachusetts Institute of Technology (MIT) Electromagnetic Equation Propagation) \cite{OskooiRo10} FDTD solver used in this work can simulate either nonlinear or absorbing medium.
The value of second-order nonlinear refractive index $n_{2\,\mathrm{in}}= 6.5\times10^{-17}$~m$^2$/W based on the measured in Ref.~\cite{Kuhl85} in the visible range third-order optical susceptibility  $\chi^{(3)}_{\mathrm{in}}\approx 2 \times 10^{-10}$~esu was utilized in the present work.
Each point in all subsequent plots refers to a separate configuration of nanoparticles on a plane.

As shown in Ref.~\cite{Panov19a}, the densely packed bidisperse monolayer of GaP spheres with the radii $r$ of 77 and 101~nm exhibits the negative refraction at $\lambda=532$~nm. 
These sizes lie just above the dipole magnetic ($r=76$~nm) and electric ($r=100$~nm) Mie resonances whose values are calculated using Debye formulas \cite{Debye1909,Hulst81}, 
\begin{align*}
\frac{2\pi r n_\mathrm{in}}{\lambda}&=c_{j-1},&&\textrm{(magnetic)}&\\
\frac{2\pi r n_\mathrm{in}}{\lambda}&=c_j \left[ 1 - \frac{1}{n_\mathrm{in}^2 j}\right] ,&&\textrm{(electric)}
\end{align*}
where $n_\mathrm{in}$ is the refractive index inside the spheres, $j=1,2,\ldots$, $c_0=\pi$, $c_1=4.493$.
The computations in Ref.~\cite{Panov19a} were conducted for the finite extinction coefficient of GaP ($\kappa=0.0026$). 
This study neglecting the extinction coefficient show the similar result for the monolayer of 77 and 101~nm radius particles, the influence of $\kappa$ on the linear optical properties of the GaP metasurface at $\lambda=532$~nm is below the standard deviation value for the computed value of the index of refraction.
The effective linear refractive index $n_{0\,\mathrm{eff}}$ of the structure with volume fraction $f=25.7$~\% is calculated neglecting $\kappa$ as $-0.39\pm0.02$. 
This value is close to one computed in Ref.~\cite{Panov19a}: $n_{0\,\mathrm{eff}}=-0.40\pm0.02$.
Here, $f$ is expressed as ratio of the net volume of the spheres to the volume of the smallest rectangular parallelepiped containing these particles.

% \section{Results and discussion}

At first, it is interesting to estimate the effective second-order nonlinear refractive index of the metasurface with negative refraction index.
The evaluation of the effective second-order nonlinear $n_{2\,\mathrm{eff}}$ refractive index for the bidisperse monolayer of the spheres with $f=25.7$~\%, $r_1=77$ and $r_2=101$~nm is $(6.5\pm0.8)\times10^{-15}$~m$^2$/W which is two orders of magnitude larger than that of the bulk gallium phosphide.
For purposes of comparison, the bidisperse monolayer consisting of spheres of artificial material with $n_0$ of GaP and  $n_2$ of gallium phosphide with negative sign was modeled: the resulting value of $n_{2\,\mathrm{eff}}=-(6.5\pm0.7)\times10^{-17}$~m$^2$/W, that is the negative index  metasurface does not show the inversion of $n_{2\,\mathrm{eff}}$ sign. It is appropriate at this point to recall that in Ref.~\cite{Panov19} the monodisperse monolayers of the spheres exhibit the inversion of effective second-order refractive index near the Mie resonances.

\begin{figure*}
{\centering
\begin{minipage}[b]{\figsize}
\includegraphics[width=\figsize]{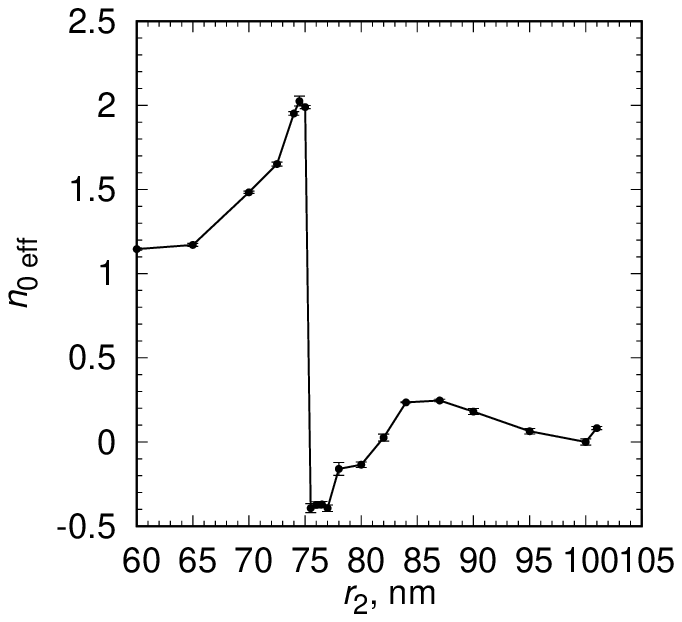}
\end{minipage}\hfill
\begin{minipage}[b]{\figsize}
\includegraphics[width=\figsize]{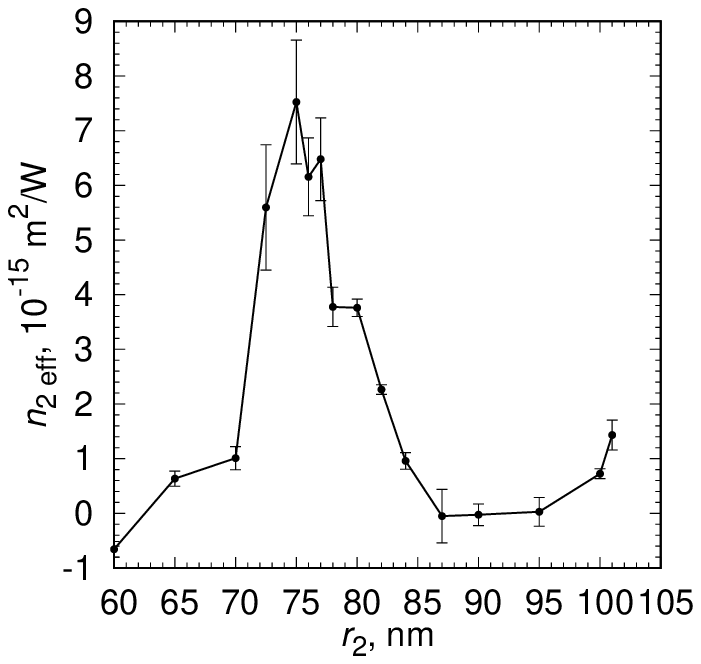}
\end{minipage}
\par} 
\caption{\label{n2_rx-101_GaP_randsurf}
Effective linear $n_{0\,\mathrm{eff}}$ and second-order nonlinear $n_{2\,\mathrm{eff}}$ refractive indices of the disordered metasurface consisting of two types of GaP spheres: with the fixed radius $r_1=101$~nm and the varying second radius $r_2$ as abscissa. The number of particles is 264 on $4\times 4$~$\mu$m area. The error bars show the standard deviations.}
\end{figure*} 

\begin{figure*}
{\centering
\begin{minipage}[b]{\figsize}
\includegraphics[width=\figsize]{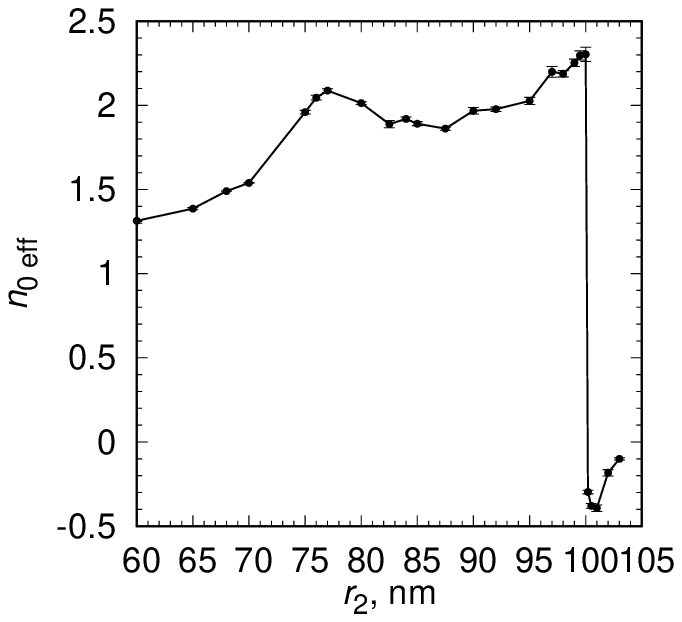}
\end{minipage}\hfill
\begin{minipage}[b]{\figsize}
\includegraphics[width=\figsize]{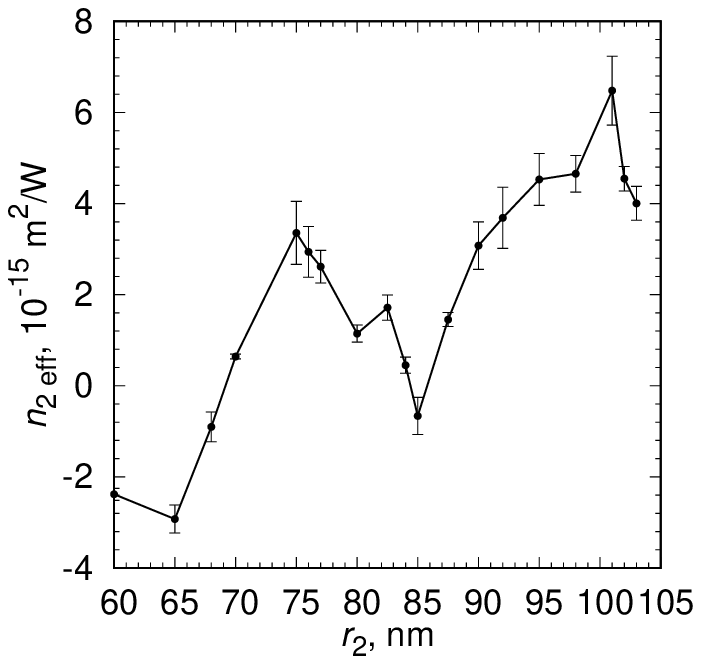}
\end{minipage}
\par} 
\caption{\label{n2_rx-77_GaP_randsurf}
Effective linear $n_{0\,\mathrm{eff}}$ and second-order nonlinear $n_{2\,\mathrm{eff}}$ refractive indices of the disordered metasurface consisting of two types of GaP spheres: with the fixed radius $r_1=77$~nm and the varying second radius $r_2$ as abscissa. The number of particles is 264 on $4\times 4$~$\mu$m area. The error bars show the standard deviations.}
\end{figure*} 

Next, the effective nonlinear refractive index of the bidisperse disordered metasurface was studied in proximity to the negative refraction condition. 
This was done as follows: the size of one half of the particles was fixed with $r_1$ and the sphere radius of another half $r_2$ was varied.
Figs.~\ref{n2_rx-101_GaP_randsurf},~\ref{n2_rx-77_GaP_randsurf} depict the results of the FDTD simulations of the metasurfaces with the sizes of the particles in the vicinity of the negative refraction state.
The behavior of $n_{0\,\mathrm{eff}}(r_2)$ for $r_1=101$~nm is  in close agreement with that obtained in Ref.~\cite{Panov19a} accounting for the absorption of GaP as long as it is sufficiently low. 
The effective linear refractive index abruptly drops and becomes negative when both $r_1$ and $r_2$ reach the sizes of the magnetic and electric dipole Mie resonances.
In a similar manner, $n_{0\,\mathrm{eff}}(r_2)$ for $r_1=77$~nm has a dip near $r_2=101$~nm where the effective linear refractive index is negative.
These dips arise from the discontinuities of effective magnetic permeability or electric permittivity at the Mie resonances (see, e.g., Refs.~\cite{Holloway03,Silveirinha11}).

The dependence of $n_{2\,\mathrm{eff}}(r_2)$ for $r_1=101$~nm (Fig.~\ref{n2_rx-101_GaP_randsurf}) begins from the negative values since the contribution of particles with sizes close to the electric dipole resonance prevails. 
This contribution is negative owing to the inversion of $n_{2\,\mathrm{eff}}$ sign for such particles \cite{Panov19}.
For a monodisperse disordered metasurface with 132 particles with $r=101$~nm arranged on $4\times 4$~$\mu$m area (one half of bidispersive metasurface) $n_{2\,\mathrm{eff}}=-(1.4\pm0.4)\times10^{-15}$~m$^2$/W.
Further, $n_{2\,\mathrm{eff}}(r_2)$ has a maximum when the metasurface combines the spheres with the electric and magnetic dipole Mie resonances.
Under these circumstances, the bidisperse disordered metasurface switches to negative refraction.
As $r_2$ moves away from the magnetic dipole resonance, the second-order refractive index falls off.
For monodisperse metasurface with 264 particles with $r=101$~nm arranged on $4\times 4$~$\mu$m area, $n_{2\,\mathrm{eff}}$ is positive: $(1.4\pm0.3)\times10^{-15}$~m$^2$/W. 
This is due to shift of the Mie resonance in densely packed arrays of nanoparticles: the resonance occurs at larger sizes of spheres.
The same reason is responsible for a local maximum of $n_{2\,\mathrm{eff}}(r_2)$ for $r_1=77$~nm
at $r_2=77$~nm when the nanostructure is monodisperse (Fig.~\ref{n2_rx-77_GaP_randsurf}).
The magnetic dipole resonance in spheres is observed at higher values of $r_2$, so that $n_{2\,\mathrm{eff}}$ in Fig.~\ref{n2_rx-77_GaP_randsurf} has a local minimum with negative magnitude near $r_2=85$~nm.
The maximum of $n_{2\,\mathrm{eff}}$ for this dependence is observed when bidisperse metasurface consists of the spheres with the radii of $r_1=77$~nm and $r_1=101$~nm.

\begin{figure*}
{\centering
\begin{minipage}[b]{\figsize}
\includegraphics[width=\figsize]{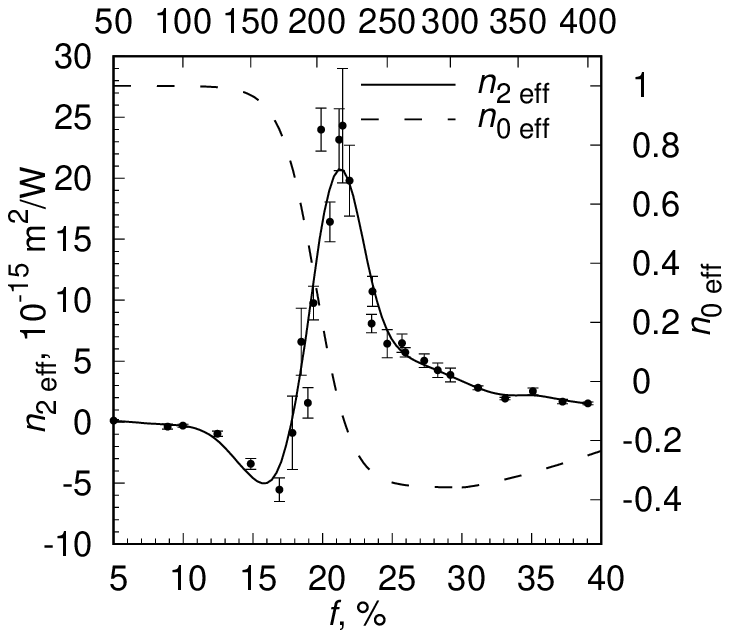}
\end{minipage}\hfill
\begin{minipage}[b]{\figsize}
\includegraphics[width=\figsize]{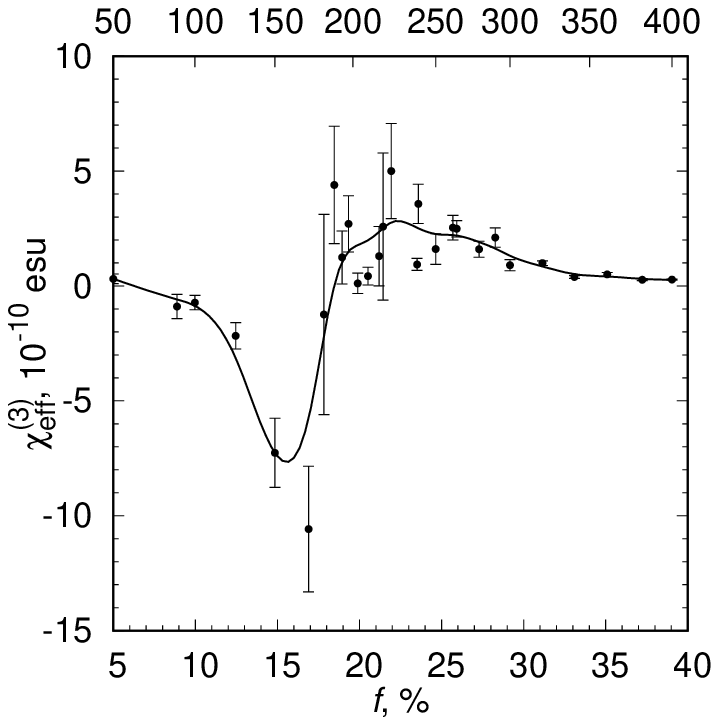}
\end{minipage}
\par}
\caption{\label{n2_conc_r77-101_GaP_surf_n2}
Effective linear $n_{0\,\mathrm{eff}}$, second-order nonlinear $n_{2\,\mathrm{eff}}$ refractive indices and third-order Kerr optical susceptibility $\chi^{(3)}_{\mathrm{eff}}$ of the disordered bidisperse metasurface consisting of equal numbers of GaP spheres with radii $r_1=77$~nm and $r_2=101$~nm as a function of volume concentration $f$. The upper abscissa axis displays the net number of the particles on $4\times 4$~$\mu$m area. The fit for effective linear refractive index $n_{0\,\mathrm{eff}}$ was taken from Ref.~\protect\cite{Panov19a}. 
The error bars show the standard deviations.}
\end{figure*} 

The next step will be to investigate the optical nonlinearity of the bidisperse metasurface during the concentration transition to the negative refraction state.
The behavior of  $n_{0\,\mathrm{eff}}$ through the transition to negative values was studied in Ref.~\cite{Panov19a}.
Fig.~\ref{n2_conc_r77-101_GaP_surf_n2} illustrates the dependencies of the effective second-order nonlinear refractive index  $n_{2\,\mathrm{eff}}$ and third-order Kerr optical susceptibility $\chi^{(3)}_{\mathrm{eff}}$ on the volume fraction of nanoparticles in the bidisperse monolayer.
At low concentrations of the GaP spheres ($f=8{-}18$~\%), the metasurface exhibits the negative value of $n_{2\,\mathrm{eff}}$ due to the inversion of optical Kerr coefficient near the electric dipole Mie resonance \cite{Panov19}.
For the higher volume fractions of nanoparticles, $n_{2\,\mathrm{eff}}$ becomes positive and has a peak when $n_{0\,\mathrm{eff}}$ crosses zero.
Further, the effective second-order nonlinear refractive index gradually decreases for the concentrations of spheres corresponding to the negative refraction.
For comparison, the third-order Kerr optical susceptibility calculated from the
values of $n_{2\,\mathrm{eff}}$ using 
% \[
\begin{equation}
 \chi^{(3)}_{\mathrm{eff}}=\frac{n_{0\,\mathrm{eff}}^2 c}{12\pi^2}n_{2\,\mathrm{eff}}
 \label{chi2n2connection}
\end{equation}
% \] 
is depicted in Fig.~\ref{n2_conc_r77-101_GaP_surf_n2}. 
It should be noted that the standard deviation of $\chi^{(3)}_{\mathrm{eff}}$ is larger than that of $n_{2\,\mathrm{eff}}$ as the uncertainty of $n_{0\,\mathrm{eff}}$ is involved.
This dependence on $f$
does not display a peak in the region of $n_{0\,\mathrm{eff}}$ close to zero.
The similar maximum of $n_{2}$ was observed in Refs.~\cite{Caspani16,Carnemolla18} for Al-doped ZnO thin films which possess epsilon-near-zero conditions at infrared wavelengths.
The authors of Refs.~\cite{Caspani16,Carnemolla18} attributed this phenomenon to the presence of the linear refractive index in a denominator of an expression for the transfer from $\chi^{(3)}$ to $n_2$.
It is worth mentioning that Eq.~\ref{chi2n2connection} or a similar formula for absorbing material \cite{delCoso04} should be carefully checked for validity for single or double negative media.

At volume fraction above 30\%, the spheres in the monolayer are densely packed and partially overlap. As a result, the resonance sizes are shifted in strongly interacting arrays of particles leading to decrease in the effective second-order nonlinear refractive index.
By this means, the disordered monolayer bidisperse metasurface has highest magnitudes of $n_{2\,\mathrm{eff}}$ under the conditions of zero index medium.

\begin{figure*}
{\centering
\begin{minipage}[b]{\figsize}
\includegraphics[width=\figsize]{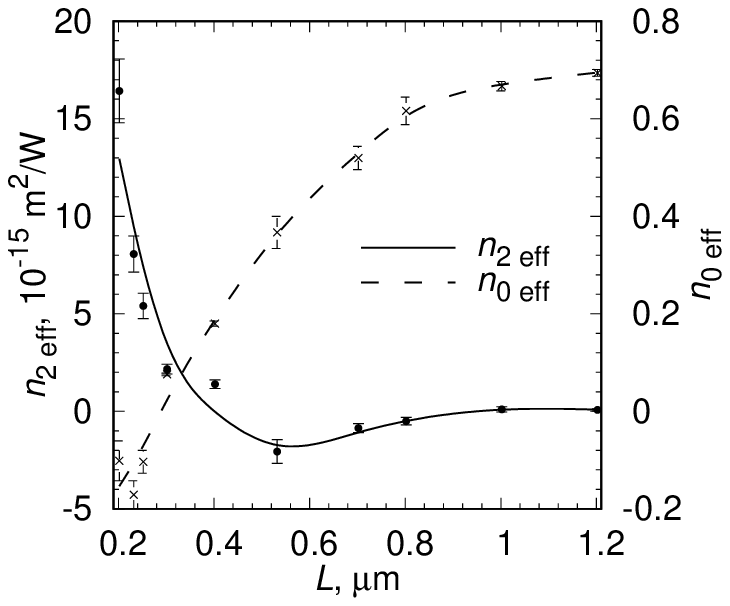}
\end{minipage}\hfill
\begin{minipage}[b]{\figsize}
\includegraphics[width=\figsize]{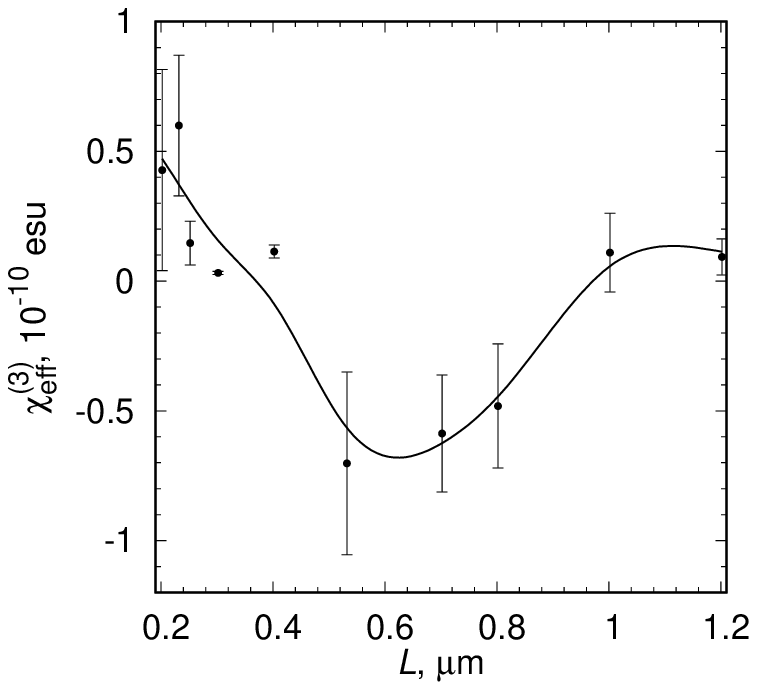}
\end{minipage}
\par}
\caption{\label{n2_ML_r77-101_GaP_surf_f20.5}
Effective linear $n_{0\,\mathrm{eff}}$, second-order nonlinear $n_{2\,\mathrm{eff}}$ refractive indices, and third-order Kerr optical susceptibility $\chi^{(3)}_{\mathrm{eff}}$ of the disordered bidisperse metasurface consisting of equal numbers of GaP spheres with radii $r_1=77$~nm and $r_2=101$~nm as  functions of the metasurface thickness $L$ at the fixed value of volume concentration $f=20.5$~\%. 
The error bars show the standard deviations.}
\end{figure*} 

Fig.~\ref{n2_ML_r77-101_GaP_surf_f20.5} demonstrates the dependence of the effective nonlinear refractive index and third-order Kerr optical susceptibility of the random bidisperse nanocomposite on its thickness $L$ for the fixed value of $f=20.5$~\%. 
The thinnest nanocomposite with $L\approx 0.2$~$\mu$m represents the monolayer used above. 
The third-order Kerr optical susceptibility $\chi^{(3)}_{\mathrm{eff}}$ is calculated using Eq.~\ref{chi2n2connection}.
For the thicknesses close to a monolayer, the metasurface possesses the condition of negative refraction.
The second-order nonlinear refractive index exhibits the largest values having the positive sign.
For larger thicknesses, the nanocomposite does not fulfill the condition of double negative medium and the sign of $n_{2\,\mathrm{eff}}$ becomes negative.
The nanocomposite is believed to be the single negative medium which displays the inversion of the $n_{2\,\mathrm{eff}}$ sign \cite{Panov19}.
The region with $n_{0\,\mathrm{eff}}\approx 0$ falls between two above thickness ranges and $n_{2\,\mathrm{eff}}$ does not show a maximum here.
For the thicker nanocomposites, $n_{0\,\mathrm{eff}}$ tends to unity and $n_{2\,\mathrm{eff}}$ is again positive with lower values on the order of $10^{-16}$~m$^2$/W.
Thus, the monolayer bidisperse metasurface shows the highest magnitudes of the second-order nonlinear refractive index.

% \section{Summary}

In conclusion, the nonlinear optical Kerr effect of bidisperse disordered monolayer nanocomposites of GaP spheres is studied  numerically.
It is displayed that this nanostructure possesses the maximum value of the second-order refractive index when the metasurface represents the mixture of nanoparticles with sizes just above the electric and magnetic dipole Mie resonances which results in the negative effective refractive index.
The effective Kerr coefficient of the negative index all-dielectric metasurface has the same sign as that of the nanoparticle material.
The bidisperse monolayer metasurfaces with different values of GaP sphere concentration exhibit a peak of the nonlinear optical Kerr effect when the effective linear index of refraction is close to zero. 
The monolayer metasurface shows the highest magnitudes of the second-order nonlinear refractive index as compared to that of thicker bidisperse nanocomposites.

% \section*{Acknowledgments}
% \subsection{Acknowledgments}
% \begin{acknowledgement}
The results were obtained with the use of IACP FEB RAS Shared Resource Center ``Far Eastern Computing Resource'' equipment (https://www.cc.dvo.ru). 
% \end{acknowledgement}

% \subsection*{Key words:} optical Kerr nonlinearity, disordered metasurface, dielectric nanoparticles, negative refraction

\bibliography{nlphase}
% \printbibliography

\end{document}